# A Reliable and Fault-Tolerant Routing for Optical WDM Networks

G.Ramesh
Department of Information Technology
KLN College of Engineering, Madurai, India
grameshphd@gmail.com

S.SundaraVadivelu
Department of Electronics and Communication Engineering
SSN College of Engineering, Kalavakkam,
Chennai, India.

*Abstract*— In optical WDM networks, since each lightpath can carry a huge mount of traffic, failures may seriously damage the end-user applications. Hence fault-tolerance becomes an important issue on these networks. The light path which carries traffic during normal operation is called as primary path. The traffic is rerouted on a backup path in case of a failure. In this paper we propose to design a reliable and fault-tolerant routing algorithm for establishing primary and backup paths. In order to establish the primary path, this algorithm uses load balancing in which link cost metrics are estimated based on the current load of the links. In backup path setup, the source calculates the blocking probability through the received feedback from the destination by sending a small fraction of probe packets along the existing paths. It then selects the optimal light path with the lowest blocking probability. Based on the simulation results, we show that the reliable and fault tolerant routing algorithm reduces the blocking probability and latency while increasing the throughput and channel utilization.

I. INTRODUCTION

*A. Wavelength-Division-Multiplexing (WDM) Networks*

The Simultaneous transmission of multiple streams of data [1] with the help of the exclusive properties of fiber optics is called as wavelength division multiplexing (WDM). The users have the capability to transfer huge amount of data at high speeds over large distances which is offered by the WDM networks.

For the backbone of future next-generation internet, wavelength division multiplexing (WDM) is considered as the most capable technology [2]. Data's are routed through optical channels called light paths, in WDM all-optical networks. The establishment of light path requires the same wavelength to be used along entire route of the light path without the wavelength conversion capability. This is commonly referred as the wavelength continuity constraint.

By allowing many independent signals with different wavelength to be transmitted simultaneously on one fiber, WDM enables the employment of a substantial portion of the available fiber bandwidth [3]. Since the wavelength determines the communication path by acting as the signature address of the origin, destination or routing, the routing and detection of these signals can be achieved independently.

Therefore the wavelength selective components are required, allowing for the transmission, recovery or routing of specific wavelengths.

*B. Fault Tolerant in WDM Networks*

Since each lightpath can carry a huge mount of traffic, failures in such networks may seriously damage end-user applications. According to the scale of their effect, failures in all-optical WDM networks can be classified into two categories [4]. One category is a wavelength-level failure which impacts the quality of transmission of each individual lightpath. The other category is a fiber-level failure which affects all the lightpaths on an individual fiber. Since each lightpath is expected to operate at a rate of several gigabytes per second, a failure can lead to a severe data loss.

The ability of network to with-stand failures is called as fault-tolerance. Failures arise due to the node failure or link failure. When a link fails all its constituent fibers also fails. All the connections which use these fibers are to be rerouted and a wavelength will be assigned. The light path which carries traffic during normal operation is called as primary path. The traffic is rerouted on a backup path in case of a failure. Optical networks which use the wavelength division multiplexing (WDM) and wavelength routing are subjected to failures. Fault tolerance becomes an important issue because of the large amount of traffic on these networks in contradiction to the conventional copper links.

Fault tolerance schemes can be broadly classified into

- Path Protection
- Restoration

*1) Path protection:* In path protection, backup resources are reserved during connection setup and both primary and backup lightpath are computed before a failure occurs. There are two types of protection schemes: dedicated and shared protection.

**Dedicated-path protection:** In dedicated-path protection (also called 1:1 protection), the resources along a backup path are dedicated for only one connection and are not shared with the backup paths for other connections.

**Shared-path protection:** In shared-path protection, the resources along a backup path may be shared with other backup paths. As a result, backup channels are multiplexed among different failure scenarios, and therefore, shared-path protection is more capacity efficient when compared with dedicated-path protection.

*2) Path Restoration:* In path restoration, the source and destination nodes of each connection traversing the failed link participate in a distributed algorithm to dynamically discover an end-to-end backup route. If no routes are available for a broken connection, then the connection is dropped.

## II. RELATED WORK

Michael T. Frederick and Arun K. Somani [6] have presented an L+1 fault tolerance which is used for the recovery of optical networks from single link failures without the allocation of valuable system resources. While the approach in its simplest form performs well against the existing schemes, the flexibility of L+1 leave many options to examine possible ways to further increase performance.

Muriel M'edard [7] has described that the protection routes are pre-computed at a single location and thus it is centralized. Before the restoration of the traffic, some distributed reconfiguration of optical switches is essential. On the other hand, restoration techniques depend upon distributed signaling between nodes or on the allocation of a new path by a central manager.

Hongsik Choi, Suresh Subramaniam and Hyeong-Ah Choi [8] have considered the network survivability which is a critical requirement in the high-speed optical networks. A failure model is considered so that any two links in the network may fail in an random order. They have presented three loop back methods of recovering from double-link failures. Only the first two methods require the identification of the failed links. But pre-computing the backup paths for the third method is more complex than the first two methods. The double link failures are tolerated by the heuristic algorithm which pre-computes the backup paths for links.

Yufeng Xin, Jing Teng, Gigi Karmous-Edwards, GeorgeN.Rouskas and Daniel [9] have studied the important fault management issue which concentrates on the fast restoration mechanisms for Optical Burst Switched (OBS) networks. The OBS network operates under the JIT signaling protocol. The basic routing mechanism is similar to the IP networks, where every OBS node maintains a local forwarding table. The entries in the forwarding table consist of the next hop information for the bursts per destination and per FEC (Forward Equivalent Class). Based on looking up the next-hop information in their forwarding tables, OBS nodes forwards the coming burst control packets and set up the connections. The connection set up process is signified by the burst forwarding or burst routing.

Jian Wang, Laxman Sahasrabuddhe and Biswanath Mukherjee [10] have considered the fault-monitoring functions which are usually provided by the optical-transmission systems. In order to measure the bit error rate in the wavelength channels using SONET framing, the B1 bit in the SONET overhead can be used. Moreover, to detect certain failures like fiber cut in other formatted optical channels, the optical power loss can be used. Optical-Electrical-Optical (OEO) conversion is used before each OXC port because most of the OXCs use electronic switching fabric. Therefore, faults can be detected on link-by-link basis. Both the end nodes of the failed link can detect the fiber cut for all-optical switches.

Lei Guo [11] has studied the problem of multiple failures in WDM networks. In order to improve the survivable performance he proposed a heuristic algorithm called Shared Multi-sub backup paths Reprovisioning (SMR). The survivable performance of SMR in multiple failures was considerably improved when compared with the previous algorithm.

Guido Maier, Achille Pattavina, Luigi Barbato, Francesca Cecini and Mario Martinelli [12] have investigated the issue of dynamic connections in WDM networks. It is also loaded with the high-priority protected static connections. They have compared various routing strategies by discrete event simulation in terms of blocking probability. Based on the occupancy cost function they have proposed a heuristic algorithm which takes several possible causes of blocking into account. The behavior of their algorithm was tested in well known case study of mesh networks, with and without wavelength conversion.

A. Rajkumar and N.S.Murthy Sharma [13] have proposed a distributed priority based routing algorithm. In order to establish the primary and backup light paths they have proposed a variety of traffic classes which uses the concept of load balancing. Based on the load on the links, their algorithm estimates the cost metric. The routing of high priority traffic was performed over the lightly loaded links. Therefore while routing the primary and backup paths, the lightly loaded links are chosen instead of choosing the links with heavier loads. The load balancing will not reflect the dynamic load changes because it is used in the routing metric.

Dong–won shin, Edwin K.P.Chong and Howard Jay Siegel [14] have developed two heuristic multipath routing schemes for survivable multipath problem called CPMR (Conditional Penalization Multipath Routing) and SPMR (Successive Penalization Multipath Routing). Their schemes use "link penalization" methods to control (but not prohibit) link-sharing to deal with the difficulties caused by the link sharing. When compared with the routing scheme that searches for disjoint paths, their methods have considerably higher routing success rates which are shown through the simulation results.

All the above existing works, did not provide the solutions based on the changing traffic load and the blocking probabilities of the paths.

## III. PROBLEM DEFINITION AND SOLUTION

Accepting as many demands as possible under the network resource constraint is the main objective of the dynamic routing algorithm. This goal can be achieved by centralized algorithm (CA) [5] through finding a primary/backup lightpath pair which uses the minimum number of free channels for the current demand. Therefore more network resources are left for the future demands. On the other hand, CA is not scalable to large networks because of its centralized nature and the failure of the Network Management System (NMS) can bring down the entire network. In contrast to this, Simple Distributed Algorithm (SDA) does not have the scalability problem and the single point of failure problem of CA. Moreover its performance in terms of the number of demand blocking is much worse than CA.

Increasing total bandwidth from source to destination is the traditional use of multiple paths to distribute data. One might anticipate that the optimal solution is caused by balancing the load among multiple paths and these approaches are proposed in IP networks. In such cases, the load for a given source-destination pair is distributed on each path in proportion to the available bottleneck bandwidth of that path. But unpredictably, in WDM networks, the above strategies lead to the worst performance. It will be better if we deliver the entire load on a single optimal path depending upon the current network-wide load status.

If the traffic load for each source and destination pair remains static, then the single static light path which is based on the load gives best performance. But it will not be optimum for the dynamic traffic load. For this case, in order to provide the choice of selecting the best light path, multiple light paths needs to be maintained based on the changing traffic load conditions.

In this paper we propose to design a reliable and fault tolerant routing (RFTR) algorithm for primary and backup paths.

In order to establish the primary path, this algorithm uses the concept of load balancing. Given a physical network with the link costs and the traffic requirements between every source-destination pair, then finding a route of the light paths for the network with least congestion, is called as .load balancing. In this algorithm, based on the load of the links the cost metric is estimated. The traffic is routed over the lightly loaded links. Therefore when routing the primary path, the links with the lighter loads are selected instead of links with the heavier loads.

Using path restoration backup paths are established. In backup path setup, the source sends a small fraction of probe packets along the existing paths. For a higher burst arrival rate, the fraction of traffic probing will be lower. For a slow changing traffic, the period of update will be higher resulting in an even smaller fraction.

The source edge can monitor and identify the requests that are rejected at the network based on receiving the PACKS/NACKS from the destinations. Thus the source can easily calculate the blocking probability through the monitored results from the probe packets. The ingress edge node selects the optimal light path with the lowest blocking probability based on the measured blocking probabilities and forwards the data through this optimal light path. On the other hand, it keeps probing the sub-optimal path for their current blocking probability.

IV. RELIABLE FAULT-TOLERANT ROUTING PROTOCOL

*A. Computing Primary Path*

The link cost function for primary path computation is designed based on the following steps:

1) For each link $L_j$, j = 1,2,3,… calculate the load index of the link $L_j$ as

$$\text{Load (LI)} = C_f / C_n \quad (1)$$

Where $C_f$ gives the number of free channels in that link and $C_n$ is the total no. of channels in that link.

2) The link cost function Cost $(L_j)$ is then defined as
Cost $(L_j)$ = 1- Load (LI), if Load (LI) > LT
= 1+ Load (LI), if Load (LI) > 0
and Load (LI) < = LT

$$= \infty, \text{ if Load (LI)} = 0 \quad (2)$$

where LT is the load threshold.

3) After we assign each link a cost using the above formula, Dijkstra's shortest path algorithm is then used to compute the least-cost path as the primary path. If the least-cost path has a cost of infinity, then the demand is blocked; otherwise a backup path is computed using the method given in the next subsection.

*B. Computing Backup Path*

Let the number of paths between source S and destination D are n.

We propose an adaptive multipath protocol to select the optimum path based on the blocking probabilities of the paths. In our approach, small fractions of probe packets are sent by the non-optimal paths such that these paths are selected very rarely. The probe packets contain sequence numbers to identify the packets.

(i) Let $P_j$, j=1,2….k be the set of probe packets sent on the paths $R_j$, j=1,2…k.

(ii) On receiving the probes packets, the destination D for the path $R_j$, send an PACK packet to the source, for each packet correctly received. The missing or dropped packets can be identified using the sequence numbers of the received packets. For each dropped packet, it sends a NACK packet to the source.

(iii) For the path $R_j$, the source calculates the blocking probability $BP_j$ such that

$$BP_j = P_{lost} / P_{sent} \quad (3)$$

Where $P_{lost}$ is the number of packets dropped and $P_{sent}$ is the number of packets sent.

(iv) Similarly for all the paths $R_j$, the source S calculates their blocking probabilities $BP_j$ based on the PACK and NACK feedback from the destination D.

(v) Now sort the paths { $R_j$, j=1,2.,…k} in ascending order of $BP_j$ values.

(vi) The paths which are having less blocking probabilities $BP_1$, $BP_2$, $BP_3$ … are selected as backup paths. If there is any

sudden or unexpected failure occurs in the primary path, traffic can be rerouted through these backup paths.

At the same time, for their blocking probability it keeps searching the sub optimal paths. Because of this, we can able to jump quickly to a new path when blocking probability of the current path increases. This occurrence is quite obvious in IP networks where the traffic patterns may vary significantly.

By sending a small fraction of traffic for probing, the aggregated throughput is reduced. However by finding a new optimal path quickly this reduction is compensated. The value of small fraction depends upon the sample size for accurately calculating the blocking probability. For a higher burst arrival rate the fraction of traffic for probing is very low.

## V. SIMULATION RESULTS

### A. Simulation Model and Parameters

In this section, we examine the performance of our reliable and fault tolerant routing algorithm (RFTR) with an extensive simulation study based upon the ns-2 network simulator [15]. We use the Optical WDM network simulator (OWNs) patch in ns2. In our simulation, we simulate an 8-Node topology as an example (Figure.1) which can be extended to any number of nodes. Various simulation parameters are given in table 1.

TABLE I. SIMULATION PARAMETERS

| Topology | Mesh |
|---|---|
| Total no. of nodes | 8 |
| Link Wavelength Number | 8 |
| Link Delay | 10ms |
| Wavelength Conversion Factor | 1 |
| Wavelength Conversion Distance | 8 |
| Wavelength Conversion Time | 0.024 |
| Link Utilization sample Interval | 0.5 |
| Traffic Arrival Rate | 0.5 |
| Traffic Holding Time | 0.2 |
| Packet Size | 200 |
| No. of Session-traffics | 4 |
| Max Requests Number | 50 |

In our experiments, we use a dynamic traffic model in which connection requests arrive at the network according to an exponential process with an arrival rate r (call/seconds). The session holding time is exponentially distributed with mean holding time s (seconds). The connection requests are distributed randomly on all the network nodes. In all the experiments, we compare the results of RFTR with DPBR [13] scheme.

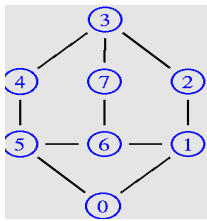

Figure: 1 8-Node Topology

### B. Performance Metrics

We measure the following metrics in all the simulation experiments:

- ➢ Blocking Probability
- ➢ Throughput in terms of Packets Received
- ➢ End-to-End Delay
- ➢ Channel Utilization

### C. Results

A. Based On Rate

In the initial experiment, we vary the traffic rate as 2Mb, 4Mb….8Mb and measure the blocking probability, end-to-end delay and channel utilization.

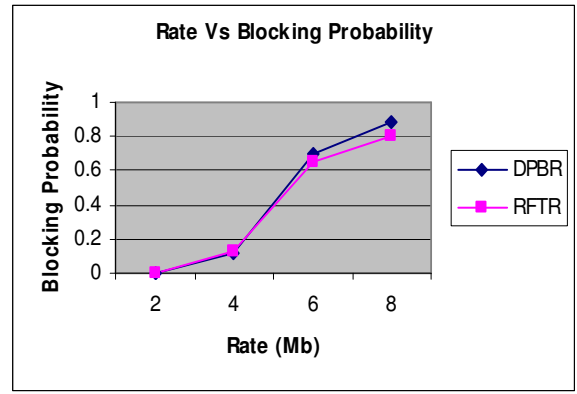

Figure: 2 Rate Vs Blocking Probability

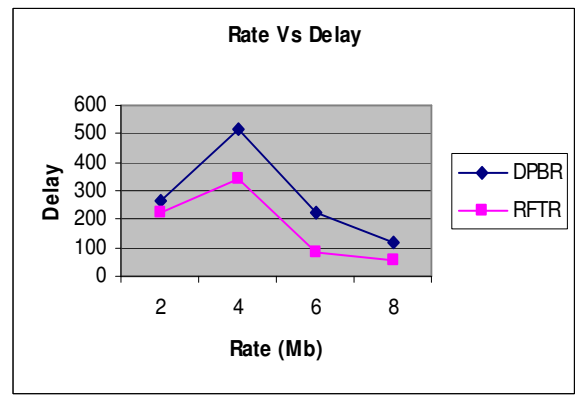

Figure: 3 Rate Vs Delay

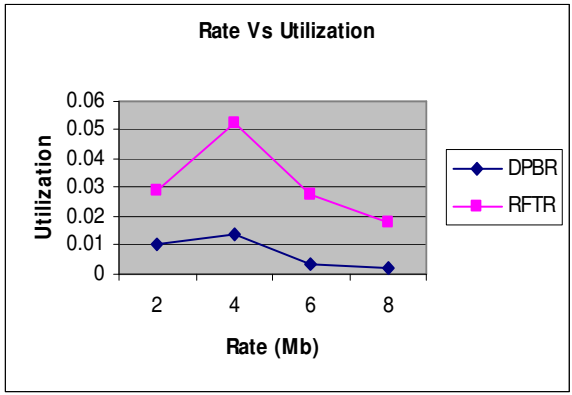

Figure: 4 Rate Vs Utilization

Figure.2 shows the blocking probability obtained with our RFTR algorithm compared with DPBR scheme. It shows that the blocking probability is significantly less than the DPBR, as rate increases.

Figure.3 shows the end-to-end delay occurred for various rates. It shows that the delay of RFTR is significantly less than the DPBR.

Figure.4 shows the channel utilization obtained for various rate. It shows that RTFR has better utilization than the DPBR scheme.

B. Based On Time

In the second experiment, we vary the time interval as 2, 4, 6….40 seconds and measure the blocking probability, end-to-end delay, throughput and channel utilization.

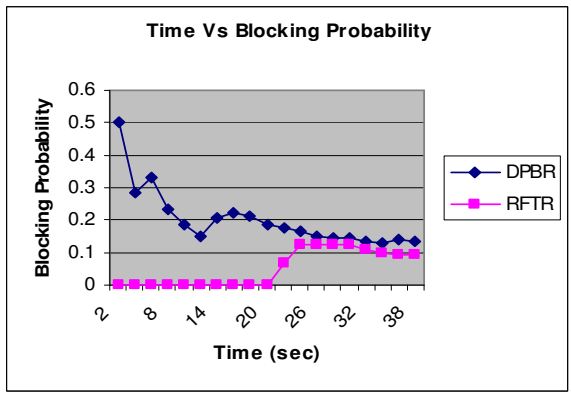

Figure: 5 Time Vs Blocking Probability

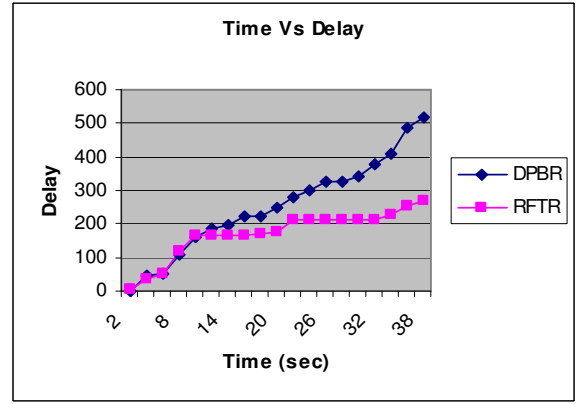

Figure: 6 Time Vs Delay

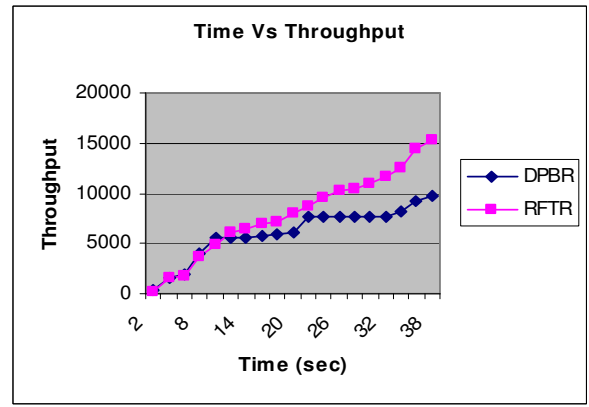

Figure: 7 Time Vs Throughput

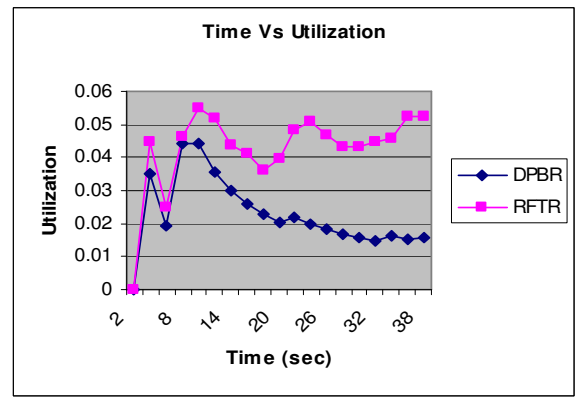

Figure: 8 Time Vs Utilization

Figure.5 shows the blocking probability obtained with our RFTR algorithm compared with DPBR scheme. It shows that the blocking probability is significantly less than the DPBR, as time increases.

Figure.6 shows the end-to-end delay occurred for various time. It shows that the delay of RFTR is significantly less than the DPBR.

Figure.7 shows the throughput occurred for various time. As we can see from the figure, the throughput is more in the case of RFTR when compared to DPBR.

Figure.8 shows the channel utilization obtained for various time. RFTR shows better utilization than the DPBR scheme.

C. Based on Traffic Sources

In this experiment, we vary the number of traffic sources as 1, 2, 3, and 4 and measure the blocking probability, end-to-end delay and throughput.

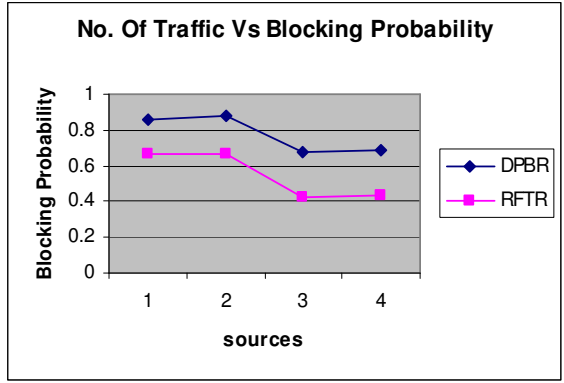

Figure: 9 Traffic Vs Blocking Probability

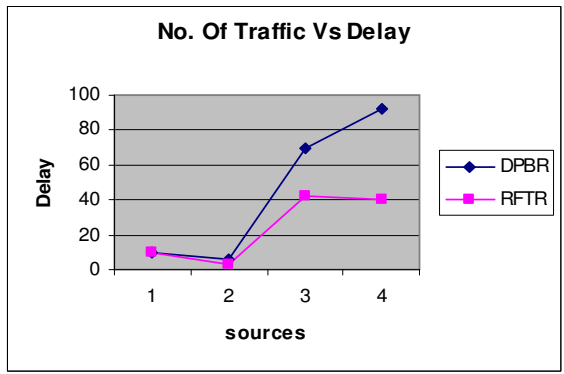

Figure: 10 Traffic Vs Delay

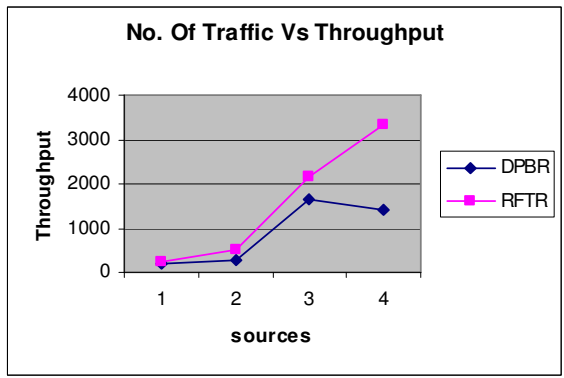

Figure: 11 Traffic Vs Throughput

Figure.9 shows the blocking probability obtained with our RFTR algorithm compared with DPBR scheme. It shows that the blocking probability of RFTR is significantly less than the DPBR, as the number of traffic sources increases

Figure.10 shows the end-to-end delay occurred when varying the number of traffic sources. It shows that the delay of RFTR is significantly less than the DPBR.

Figure.11 shows the throughput occurred when varying the number of traffic sources. As we can see from the figure, the throughput is more in the case of RFTR when compared to DPBR.

VI. CONCLUSION

In this paper we have designed a reliable and fault-tolerant routing algorithm for establishing primary and backup paths in optical WDM networks. In order to establish the primary path, this algorithm uses load balancing in which link cost metrics are estimated based on the current load of the links. The traffic is routed over the lightly loaded links. Therefore the links with the lighter loads are selected instead of links with the heavier loads. In backup path setup, the source sends a small fraction of probe packets along the existing paths. It can monitor and identify the requests that are rejected at the network based on the received positive and negative feedback from the destinations. The source then calculates the blocking probability from the received feedback and selects the optimal light path with the lowest blocking probability. Based on the simulation results, we have shown that the reliable and fault tolerant routing algorithm reduces the blocking probability and latency while increasing the throughput and channel utilization.

As a future work, we will concentrate on designing an efficient fault detection and localization technique in WDM networks.

REFERENCES

[1] Canhui (Sam) Ou Hui Zang ,Narendra K. Singhal, Keyao Zhu, Laxman H. Sahasrabuddhe, Robert A. Macdonald, And Biswanath Mukherjee, "Sub path Protection For Scalability And Fast Recovery In Optical WDM Mesh Networks", IEEE Journal On Selected Areas In Communications, Vol. 22, No. 9, November 2004.

[2] Vinh Trong Le, Son Hong Ngo, Xiao Hong Jiang, Susumu Horiguchi and Yasushi Inoguchi, "A Hybrid Algorithm for Dynamic Lightpath Protection in Survivable WDM Optical Networks", IEEE,2005

[3] Sateesh Chandra Shekhar, "Survivable Multicasting in WDM Optical Networks", August,2004

[4] S.Ramamurthy, Laxman Sahasrabuddhe, and Biswanath Mukherjee, "Survivable WDM Mesh Networks", Journal of light wave technology, vol. 21, no. 4, April 2003.

[5] Lu Ruan, Haibo Luo, and Chang Liu, "A Dynamic Routing Algorithm with Load Balancing Heuristics for Restorable Connections in WDM Networks", IEEE Journal on Selected Areas in Communications, Vol. 22, no. 9, November 2004.

[6] Michael T. Frederick and Arun K. Somani, "A single-fault recovery strategy for optical Networks using sub graph routing", The International Journal of Computer and Telecommunications Networking, Volume 50, Issue 2 (February 2006)

[7] MurielM´edard, "Network Reliability and Fault Tolerance", IEEE Transactions on Volume 50, Issue 1, Page(s):85 – 91, Mar 2001

[8] Hongsik Choi, Suresh Subramaniam, and Hyeong-Ah Choi, "On Double-Link Failure Recovery in WDM Optical Networks", INFOCOM Volume: 2, On page(s): 808- 816 vol.2, 2002

[9] Yufeng Xin, Jing Teng, Gigi Karmous-Edwards,GeorgeN.Rouskas and Daniel Stevenson, "Fault Management with Fast Restoration for Optical Burst Switched Networks", IEEE,2004

[10] Jian Wang, Laxman Sahasrabuddhe and Biswanath Mukherjee, **"Fault Monitoring and Restoration in Optical WDM Networks"**, National Fiber Optic Engineers Conference NFOEC 2002

[11] Lei Guo, "Heuristic Survivable Routing Algorithm for Multiple Failures in WDM Networks", in proc. of 2nd IEEE/IFIP International Workshop on Broadband Convergence Networks, pp: 1-5, 21 May 2007, Doi: 10.1109/BCN.2007.372750.

[12] Guido Maier, Achille Pattavina, Luigi Barbato, Francesca Cecini and Mario Martinelli, "Routing Algorithms in WDM Networks under Mixed Static and Dynamic Lambda-Traffic", Journal on Photonic Network Communications, vol. 8, no. 1, pp: 69- 87, June 2004, Doi: 10.1023/B:PNET.0000031619.18955.b4.

[13] Rajkumar and Dr.N.S.Murthy Sharma, "A Distributed Priority Based Routing Algorithm for Dynamic Traffic in Survivable WDM Networks", International Journal of Computer Science and Network Security, vol.8, no.11, November 2008.

[14] Dong-won Shin, Edwin K.P.Chong and Howard Jay Siegel, "Survivable Multipath Routing Using Link Penalization", Computing and Communications, 2004 IEEE International Conference.

[15] Network Simulator – www.isi.edu/nsnam

ABOUT AUTHORS

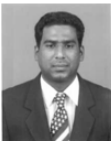

**G.Ramesh** is working as Assistant Professor in the Department of Information Technology at KLN College of Engineering, Madurai, India. He has completed the graduation in Electrical and Electronics Engineering and Post Graduation in Computer Science and Engineering. He is doing research work in optical networking

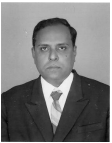

**Dr.S.SundaraVadivelu** is working as Professor in Department of Electronics and Communication Engineering at SSN College of Engineering, Kalavakkam, Chennai, India. His research interest is in Optical Communication and Networks.